\pdfoutput=1
\documentclass[%
  reprint,
  superscriptaddress,
  nofootinbib,
  amsmath,amssymb,
  aps,
  prd,
  longbibliography
]{revtex4-2}

\usepackage[utf8]{inputenc}
\usepackage{microtype}
\usepackage{lmodern}
\usepackage{siunitx}
\usepackage[colorlinks,allcolors=blue,linktocpage]{hyperref}
\usepackage{cleveref}
\crefname{equation}{Eq.}{Eqs.}
\crefname{figure}{Fig.}{Figs.}
\usepackage{tikz}
\usetikzlibrary{backgrounds}
\usetikzlibrary{intersections}
\usetikzlibrary{patterns}
\usepackage{pgfplots}
\pgfplotsset{grid style={densely dotted,gray}}
\pgfplotsset{compat=1.14}

\begin{document}
\boldmath
\title{\texorpdfstring{Projected NA62 sensitivity to heavy neutral lepton production in $K^+\to\pi^0e^+N$ decays}{Projected NA62 sensitivity to heavy neutral lepton production in K+ -> pi0 e+ N decays}}
\unboldmath

\author{Jean-Loup Tastet}
\email{jeanloup@nbi.ku.dk}
\affiliation{Niels Bohr Institute, University of Copenhagen, Blegdamsvej 17, DK-2010, Copenhagen, Denmark}

\author{Evgueni Goudzovski}
\email{eg@hep.ph.bham.ac.uk}
\affiliation{School of Physics and Astronomy, University of Birmingham, Edgbaston, Birmingham, B15 2TT, United Kingdom}

\author{Inar Timiryasov}
\email{inar.timiryasov@epfl.ch}
\affiliation{Institute of Physics, Laboratory for Particle Physics and Cosmology,
\'{E}cole Polytechnique F\'{e}d\'{e}rale de Lausanne, CH-1015 Lausanne,
Switzerland}

\author{Oleg Ruchayskiy}
\email{oleg.ruchayskiy@nbi.ku.dk}
\affiliation{Niels Bohr Institute, University of Copenhagen, Blegdamsvej 17, DK-2010, Copenhagen, Denmark}

\begin{abstract}
Heavy neutral leptons (HNLs) appear in many extensions of the Standard Model of particle physics.
In this study, we investigate to which extent the NA62 experiment at CERN could improve the existing bounds on the HNL mixing angle $|U_e|^2$ by performing a missing mass search in $K^+\to\pi^0 e^+N$ decays in flight.
We show that the limit $|U_e|^2 \simeq 2\times 10^{-6}$ can be reached with the currently available data in the mass range $125$ -- $\SI{144}{MeV}$, which is currently not well covered by production searches.
Future data, together with a dedicated trigger and/or improvements in rejection of out-of-acceptance photons, can improve this limit by another order of magnitude.
\end{abstract}

\maketitle

\providecommand{\mSqMiss}{\ensuremath{m_{\mathrm{miss}}^2}}
\providecommand{\diff}{\ensuremath{\mathrm{d}}}
\providecommand{\ie}{i.e.\@}
\providecommand{\eg}{e.g.\@}
\providecommand{\VPMNS}{\ensuremath{V^{\mathrm{\scriptscriptstyle PMNS}}}}
\renewcommand{\L}{\mathrm{L}} 
\providecommand{\R}{\mathrm{R}} 
\providecommand{\Pdecay}{\ensuremath{f_{\mathrm{decay}}^{\mathrm{vis}}}}

\section{Introduction}

\subsection{Heavy Neutral Leptons}

Despite its astounding success in describing the outcomes of collider experiments, the Standard Model of particle physics (SM) fails to account for multiple reliable observations: the baryon asymmetry of the Universe (BAU, see \eg{} Ref.~\cite{Canetti:2012zc}), dark matter (see \eg{} Ref.~\cite{Peebles:2017bzw}), as well as neutrino flavor mixing and oscillations~\cite{Bilenky:2014ema}.
The latter observations provide unambiguous evidence for non-zero neutrino masses, which call for the introduction of additional degrees of freedom into the SM.
Among many models explaining neutrino masses, those that introduce no new particles above the electroweak scale are of special interest, since they do not destabilize the Higgs mass~\cite{Vissani:1997ys,Shaposhnikov:2007nj,Bezrukov:2012sa} and are accessible already by the current generation of experiments (see \eg{} Ref.~\cite{Beacham:2019nyx}).

Such particles may appear for example in extensions of the neutrino sector (see \eg{} Refs.~\cite{Alekhin:2015byh,Strategy:2019vxc}) such as the type-I seesaw theories~\cite{Minkowski:1977sc,GellMann:1980vs,Mohapatra:1979ia,Yanagida:1980xy,Schechter:1980gr,Schechter:1981cv}.
The assignment of charges in the SM predicts that hypothetical right-handed counterparts to neutrinos would be completely neutral, \ie{} transform as singlets under the SM gauge group.
As such, they also admit a Majorana mass term whose value is not predicted from neutrino data.
The physical spectrum of these theories contains three light neutrino mass states $\nu_{\L i}$ plus a number of new \textit{heavy neutral leptons} (HNLs) $N_{\R I}$ (conventionally defined as right-handed to be consistent with other $SU(2)_L$ singlet fermions).
These heavy neutral leptons inherit from the active neutrino flavor states their weak-like interactions with $W$ and $Z$ bosons, albeit with a coupling suppressed by the (flavour-dependent) elements of the mixing matrix $\Theta_{\alpha I} \ll 1$. In what follows, we will refer to the elements of this matrix as \textit{mixing angles}.
The active neutrino flavors $\nu_{\L \alpha}$ ($\alpha=e,\mu,\tau$) are then a superposition of light and heavy mass states: $\nu_{\L \alpha} = \VPMNS_{\alpha i} \nu_{\L i} + \Theta_{\alpha I} N_{\R I}^c$, where $\VPMNS_{\alpha i}$ is the (now non-unitary) PMNS matrix (see \eg{} Ref.~\cite{Giganti:2017fhf}).

HNLs can by themselves resolve the aforementioned beyond-the-Standard-Model puzzles, as in the \textit{Neutrino Minimal Standard Model} ($\nu$MSM)~\cite{Asaka:2005an,Asaka:2005pn}.
Or they can serve as a \textit{portal} (mediator) between the SM sector and other hypothetical sectors containing new particles~(see \eg{} Refs.~\cite{delAguila:2008ir,deVries:2020qns,Liao:2016qyd,Chala:2020vqp,Li:2021tsq}
or \cite{Alekhin:2015byh,Beacham:2019nyx} for an overview).
In the latter case HNLs can possess other types of interactions (see \eg{} Refs.~\cite{Mohapatra:1979ia,Mohapatra:1980yp,Mohapatra:1986bd,Foot:1988aq,Pilaftsis:1991ug,Accomando:2016rpc,Caputo:2017pit,Basso:2008iv,Gninenko:2010pr,Batell:2016zod,Magill:2018jla,Fischer:2019fbw,Chala:2020vqp}), in addition to those inherited from their mixing with the active flavor states.

In this paper, we consider a simplified model containing one HNL $N$ with three flavour mixing angles $U_\alpha \ll 1$.
It can be thought either as a single Majorana mass state, or several HNLs degenerate in mass, in which case the equivalent mixing angle that we constrain is $|U_{\alpha}|^2 = \sum_I |\Theta_{\alpha I}|^2$.

\subsection{Missing mass searches}

Intensity frontier experiments like NA62 at CERN are, thanks to the high statistics available, well suited to constrain HNLs. There are two main experimental methods to search for HNLs: production and decay searches~\cite{Beacham:2019nyx}. Production searches consist in reconstructing the ``missing'' momentum of invisible particles from an otherwise known kinematical configuration, and searching for a mass peak emerging over a smooth background --- which indicates the presence of a new particle. They can be performed only if the kinematics of the process are fully known, as \eg{} at kaon factories or $e^+ e^-$ colliders. Decay searches consist in identifying visible final states in the HNL decays and can be performed at fixed-target, beam dump, $e^+ e^-$ or $pp$ collider based experiments. Production searches are sensitive to the HNL production rate alone, but not to its lifetime or decay modes.\footnote{Provided the lifetime is long enough that the HNL does not decay visibly within the experimental setup.} In typical models, the production rate is proportional to the square of a single mixing angle active in the production process, $|U_\alpha|^2$ ($\alpha=e$, $\mu$ or $\tau$). A non-observation can therefore be directly translated into a limit on this mixing angle, with little model dependence. On the other hand, decay searches are sensitive to a combination of the various squared mixing angles involved in the HNL production, multiplied by the partial HNL decay width\footnote{Unless the HNL decays promptly, in which case it is the branching fraction that matters, not the partial width.}, which in typical models also depends on squared mixing angles. The signal is thus proportional to a combination of fourth powers of mixing angles.
To be translated into a set of exclusion limits, a non-observation must therefore be interpreted within a specific model to disentangle the contributions of the various flavors, hence introducing additional model dependence.

\subsection{The NA62 experiment}

The NA62 experiment at CERN~\cite{NA62:2017rwk} employs a high intensity, almost monochromatic secondary $K^+$ beam of \SI{75}{GeV} momentum to measure the rate of the ultra-rare $K^+\to\pi^+\nu\bar\nu$ decay to a $10\%$ precision using the decay in flight technique~\cite{NA62:2017rwk}.
The beam is delivered into a \SI{80}{m} long vacuum tank, giving rise to a $K^+$ decay rate in the tank of about \SI{5}{MHz}. Both the incoming kaons and their visible decay products are detected, allowing to reconstruct the missing momentum. The experiment is equipped with a system of veto detectors for both charged and neutral particles. In particular, the photon veto helps reducing the contribution of undetected photons and $\pi^0$ mesons to the missing momentum. This leads to favourable background conditions, and provides sensitivity to $K^+$ decays with invisible particles in the final state, which are reconstructed using the \textit{missing mass} technique. Such searches have been performed~\cite{CortinaGil:2019nuo,NA62:2020mcv} or are planned both for HNLs and for other feebly interacting particles~\cite{Dobrich:2017yoq,Lanfranchi:2017wzl,Mermod:2017ceo,Drewes:2018gkc,Gori:2020xvq}.

The NA62 collaboration has recently performed a search for HNL ($N$) production in the $K^+\to e^+N$ decay with the full Run 1 (2016 -- 2018) data set, and established stringent limits at the level of $|U_e|^2 \sim 10^{-9}$ in the HNL mass range $144$ -- $\SI{462}{MeV}$~\cite{NA62:2020mcv}. The sensitivity of this search deteriorates abruptly at lower HNL masses due to the shape of the background.

\subsection{Other existing bounds}

Along with NA62 searches, other experiments have also probed the existence of HNLs mixing with $\nu_e$ in the mass range of interest \cite{Aguilar-Arevalo:2017vlf,Yamazaki:1984sj,Bernardi:1987ek}.
Below approximately \SI{120}{MeV}, the leading constraint (at the level of $|U_e|^2 \sim 10^{-8}$) comes from the search for the $\pi^+\to e^+N$ decay in flight at PIENU~\cite{Aguilar-Arevalo:2017vlf}, but it weakens sharply in the vicinity of the pion threshold. Above \SI{144}{MeV}, the best constraint is the one from the $K^+\to e^+N$ search at NA62~\cite{NA62:2020mcv}.
An earlier experiment at KEK \cite{Yamazaki:1984sj} was able to set limits covering the $120$ -- $\SI{144}{MeV}$ range, but these were significantly weaker, at the level of $|U_e|^2 \sim 10^{-6}$.
As a result, \emph{production} searches only weakly constrain the mass range $120$ -- $\SI{144}{MeV}$.

A notable exception is the PS191 experiment at CERN~\cite{Bernardi:1987ek}, which have reported competitive bounds in the above mass range (see Ref.~\cite{Ruchayskiy:2011aa} for a re-analysis including the neutral current contribution). The PS191 experiment was designed specifically to detect \emph{decay} products of heavy neutrinos in a low-energy neutrino beam produced by kaon and pion decays. Such a \emph{decay search} assumes that both the production and the visible decay of the HNL are determined by the mixing angles $|U_\alpha|^2$ and $|U_\beta|^2$ (for this reason it is often called a $|U|^4$ experiment, see \eg{} Refs.~\cite{Ruchayskiy:2011aa,Alekhin:2015byh}).
The sensitivity of these experiments crucially depends on the ratio $|U_e|^2:|U_{\mu}|^2:|U_{\tau}|^2$ of the mixing angles (\eg{} a large $|U_{\mu/\tau}|^2$ mixing angle can reduce the branching ratio of the sought-after decay $N \to e^+ e^- \nu_{\alpha}$ by suppressing the relative charged current contribution). Marginalizing over these unknown parameters can thus lead to a significantly weaker sensitivity, as shown explicitly in Refs.~\cite{Bondarenko:2021cpc,Tastet:2021thesis,Tastet:2021vwp}.

The missing mass searches therefore remain a viable option to explore a wide class of models.
This includes models where the HNL decays to invisible final states (``dark sectors'').
In particular, these searches are not affected by the values of the other mixing angles $|U_{\mu/\tau}|^2$, as long as those remain below the current experimental limits.
This motivates the present study, which consists in probing the $120$ -- $\SI{144}{MeV}$ mass range for the electron mixing at NA62 using the missing mass technique in the $K^+\to\pi^0 e^+ N$ channel.

\section{Signal simulation}
\label{sec:signal}

The proposed search involves the final state consisting of a positron and two photons originating from a prompt $\pi^0$ decay. The expected number of signal events is
\begin{equation}
    s_{\mathrm{tot}} = N_K \times \mathrm{BR}(K^+ \to \pi^0 e^+ N) \times \epsilon_{\mathrm{sig}} \times (1-\Pdecay{}),
\end{equation}
where $N_K$ denotes the effective number of $K^+$ decaying within the fiducial volume (as defined in \cite{NA62:2020mcv}), $\Pdecay{}$ is the fraction of HNLs which decay \emph{visibly} inside the detector (in which case the event is ignored by the present analysis), and $\epsilon_{\mathrm{sig}}$ is the signal detection efficiency (including the geometrical acceptance, but not the probability of the HNL decaying outside the detector).

$\Pdecay{}$ is a model-dependent parameter determined by the specific HNL decay channels. However, in most models of interest it turns out to be negligible. In the simplest model of HNLs (with only neutrino-like interactions), existing bounds constrain the lifetime to be many orders of magnitude larger\footnote{Saturating the existing bounds on the three mixing angles, we find for this minimal HNL model a conservative lower bound of $\sim\SI{1000}{km}$ on the HNL lifetime.} than the typical size of the NA62 detector, hence suppressing \Pdecay{}. If we consider instead an HNL that acts as a portal to a hidden sector and may decay invisibly with a possibly large width, then \Pdecay{} remains small since the invisible width suppresses the branching fraction of visible decays. Only in the case where non-minimal interactions lead to a large enhancement of those \emph{visible} decays can \Pdecay{} become large enough to suppress the signal.

\begin{figure}
    \centering
    \input{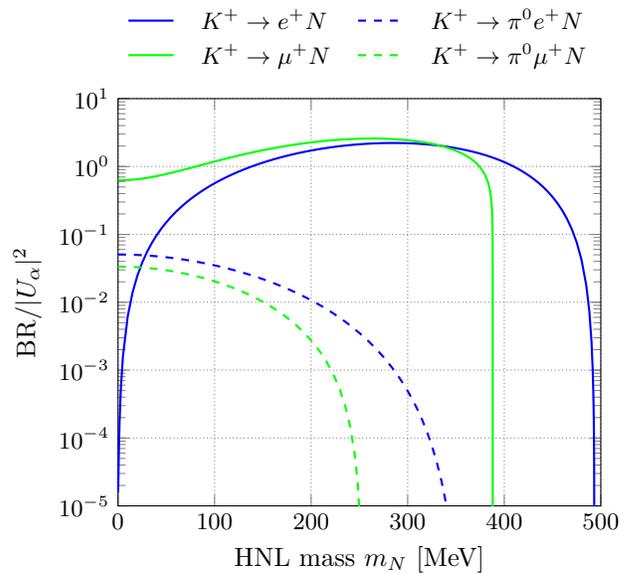}%
    \caption{HNL production branching ratios in leptonic and semileptonic $K^+$ decays, normalised to the squares of the relevant mixing angles. Contrary to the $K^+\to e^+N$ decay, the $K^+\to\pi^0 e^+ N$ decay considered in this study is not helicity-suppressed for $m_N \ll m_{K^+}$.}
    \label{fig:production_br}
\end{figure}

The matrix element of the decay, and the branching ratio $\mathrm{BR}(K^+\to\pi^0 e^+N)$, both of which depend on the assumed HNL mass $m_N$, are computed following Refs.~\cite{Gorbunov:2007ak,Bondarenko:2018ptm}, using the measured form factors from Ref.~\cite{Lazzeroni:2018glh}. The branching ratio is shown as a function of $m_N$ as the blue dashed line in \cref{fig:production_br}. The NA62 Run 1 data sample currently available for the $K^+\to\pi^0 e^+N$ search, collected using a 1-track trigger with an effective prescaling factor of about 150, corresponds to $N_K\approx 3\times 10^{10}$~\cite{NA62:2020mcv}. The acceptance $\epsilon_{\mathrm{sig}}$ is computed by interfacing our matrix element sampler with the full \textsc{Geant4}-based NA62 simulation framework~\cite{GEANT4:2002zbu}, and employing a basic event selection requiring a positron and two photons from a $\pi^0\to\gamma\gamma$ decay in the geometric acceptance of the detector. The acceptance is found to be about 10\% for HNL masses below \SI{150}{MeV}, and to decrease as a function of $m_N$ for higher masses towards the kinematic endpoint.

The events are binned in squared missing mass ${\mSqMiss = (p_{K^+} - p_{\pi^0} - p_{e^+})^2}$. The finite momentum and energy resolution of the detector causes the reconstructed signal \mSqMiss{} distribution to follow a Gaussian profile centered at $m_N^2$. The typical NA62 resolution on \mSqMiss{} is about $10^{-3}\,\si{GeV^2}$~\cite{NA62:2020mcv}. A value of $\sigma_{m^2}=\SI{1.7e-3}{GeV^2}$ obtained for the $K^+\to e^+\nu$ decay~\cite{NA62:2020mcv} is assumed conservatively for this study.

\section{Background estimate}
\label{sec:background}

\begin{figure}
    \centering
    \input{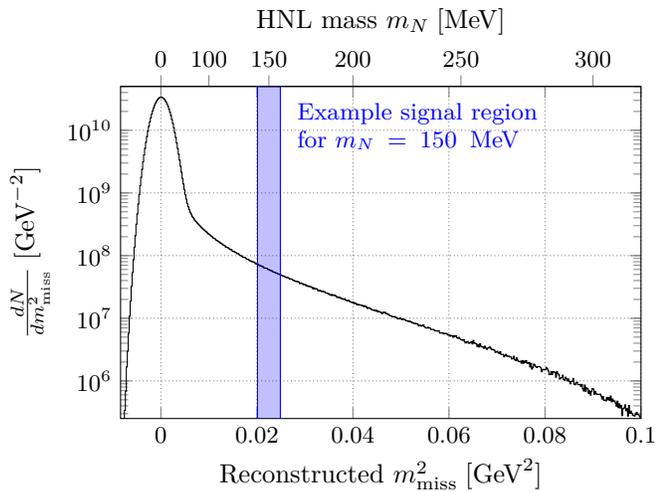}%
    \caption{Reconstructed squared missing mass spectrum of the $K^+\to \pi^0e^+\nu_e \gamma$ background, obtained by modelling the NA62 acceptance and resolution, and assuming that the radiative photon is not detected. The total number of reconstructed events in the spectrum is $1.5\times 10^8$, corresponding to $N_K=3\times 10^{10}$ kaon decays considered. A $\pm1.4\sigma_{m^2}$ wide signal region for $m_N=\SI{150}{MeV}$ is shown for illustration.}
    \label{fig:background_m2miss}
\end{figure}

The dominant source of background to the $K^+\to\pi^0 e^+N$ process comes from the radiative $K^+\to\pi^0 e^+\nu_e\gamma$ inner-bremsstrahlung decay with the radiative photon escaping detection, thus causing \mSqMiss{} to be mis-reconstructed. The expected reconstructed \mSqMiss{} spectrum of the $K^+\to\pi^0 e^+\nu_e \gamma$ process, simulated according to Ref.~\cite{Gatti:2005kw}, taking into account the NA62 acceptance and resolution, and assuming that the radiative photon is not detected, is shown in \cref{fig:background_m2miss}. The principal contribution to the background for HNL masses above \SI{100}{MeV} comes from the radiative tail, while the contributions from the main $K^+\to\pi^0e^+\nu_e$ peak at $\mSqMiss{} = 0$ (caused by the finite mass-squared resolution) and from the non-Gaussian reconstruction tails are subleading. The origin and properties of this background are similar to those encountered in the search for the $K^+\to\mu^+N$ decay at NA62~\cite{CortinaGil:2021gga}.

Other background sources, such as $K^+\to\pi^0\pi^0e^+\nu_e$ decays with both photons from a $\pi^0$ decay evading detection, or $K^+\to\pi^0\mu^+\nu_\mu$ decays followed by $\mu^+\to e^+\bar{\nu}_\mu\nu_e$ decays, are found to be subleading. In particular, the misreconstruction of the $K^+\to e^+$ decay vertex position in the latter case typically leads to the invariant mass of the two photons from the $\pi^0\to\gamma\gamma$ decay, reconstructed assuming photon emission at the decay vertex, being incompatible with the $\pi^0$ mass.

\begin{figure}
    \centering
    \input{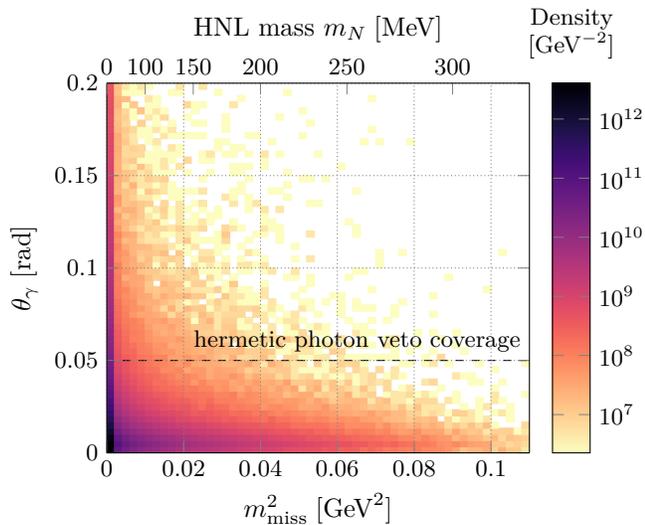}%
    \caption{$K^+\to\pi^0e^+\nu_e \gamma$ background event density as a function of the true missing mass squared and angle $\theta_\gamma$ between the photon and the beam axis. Hermetic geometric coverage is provided for photons with $\theta_\gamma<\SI{50}{mrad}$.}
    \label{fig:m2miss_vs_theta}
\end{figure}

The background from radiative photons is largely reducible thanks to the NA62 photon veto system, which provides hermetic geometric coverage for photon emission angles $\theta_\gamma$ up to $\theta_{\mathrm{max}}=\SI{50}{mrad}$ with respect to the beam axis, and partial geometric coverage (of approximately $\theta_{\mathrm{max}}/\theta_\gamma$) for larger emission angles. The nominal detection inefficiency for energetic photons (with energies in excess of a few hundred MeV) is $10^{-3}$ for the large-angle system, and well below $10^{-3}$ for the intermediate and small angles~\cite{NA62:2017rwk}. As can be seen in \cref{fig:m2miss_vs_theta}, most of the photons from $K^+\to\pi^0 e^+\nu_e \gamma$ decays susceptible to contaminate the relevant signal regions (for $m_N \gtrsim \SI{100}{MeV}$, \ie{} $\mSqMiss{} \gtrsim \SI{0.01}{GeV^2}$) are emitted within \SI{50}{mrad} of the beam axis. A simplified photon detection efficiency model is used in this study: the nominal detection inefficiency of $10^{-3}$ is assumed for $\theta_\gamma<\theta_{\mathrm{max}}$ (this assumption is valid as the energy of the photons intersecting the large-angle veto acceptance for the \mSqMiss{} range of interest is always above 200~MeV, and is typically in the GeV range), and zero detection efficiency is assumed conservatively for the (softer) photons emitted at $\theta_\gamma\ge\theta_{\mathrm{max}}$. In this model, the background events are dominated by those with soft photons emitted at angles above \SI{50}{mrad} outside the hermetic coverage zone. Therefore the accuracy of the detection efficiency model does not significantly affect the background estimate.

\section{Projected NA62 sensitivity}

\begin{figure}
    \centering
    \input{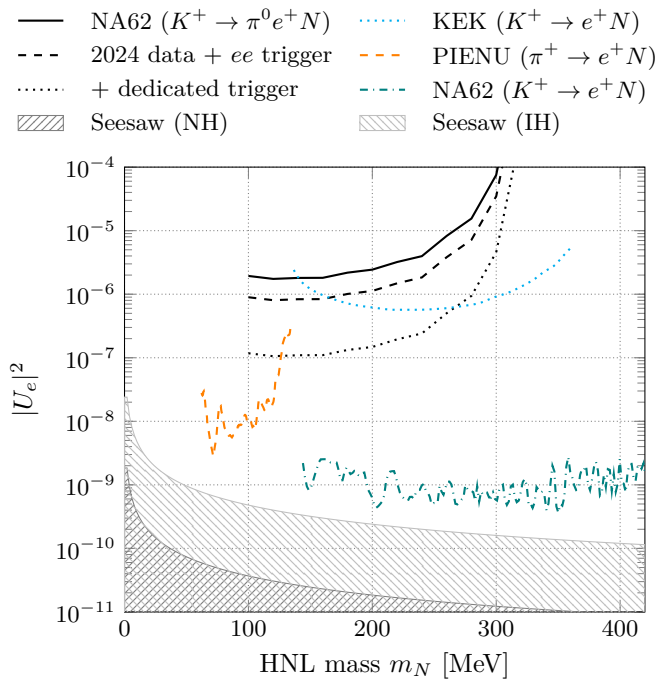}%
    \caption{Projected exclusion reach of NA62 for an HNL produced in the $K^+ \to \pi^0 e^+ N$ channel (black lines), compared to the exclusion limits set by previous missing mass searches.
    The seesaw ``bounds'' on $|U_e|^2$ are plotted under the assumption that two quasi-degenerate HNLs are fully responsible for neutrino oscillations (see the main text for details).}
    \label{fig:sensitivity}
\end{figure}

To estimate the projected sensitivity, we use for simplicity a cut-and-count analysis. We expect that the actual search will instead involve spectrum shape analysis. We define the signal region for an HNL of mass~$m_N$ as a rolling window of missing mass squared $\mSqMiss{} \in [m_N^2 - k\sigma_{m^2}, m_N^2 + k\sigma_{m^2}]$, where the width $\sigma_{m^2}= \SI{1.7e-3}{GeV^2}$ corresponds to the approximate mass-squared resolution of the detector~\cite{NA62:2020mcv} and the constant $k=1.4$ is chosen to maximize the $s/\sqrt{b}$ ratio (where $s$ and $b$ respectively denote the numbers of signal and background events inside the window) and therefore the power of the search. A typical signal region is shown in \cref{fig:background_m2miss}.
Real photon emissions produce a smoothly falling background in \mSqMiss{}. The search is performed by looking for a significant excess of events over the background count~$b$ in each signal region. The detection sensitivity is expressed as a $90\%$ confidence limit (local significance), which roughly corresponds to $s \gtrsim 1.282\sqrt{b}$ in the limit $b \gg 1$, with~$s = \mathrm{erf}(k/\sqrt{2}) \times s_{\mathrm{tot}}$ the approximate number of signal events inside the signal region. The projected, median \emph{exclusion} limit is similarly obtained, by replacing $\sqrt{b}$ with $\sqrt{b+s}$.
The background~$b$ from real photon emissions, integrated over a small \mSqMiss{} window, is approximately:
\begin{equation}
    b(\mSqMiss{}) \approx 2 k \times \sigma_{m^2} \times \langle\epsilon_{\mathrm{bkg}}\rangle \times \frac{\diff N(K^+ \to \pi^0 e^+ \nu_e \gamma)}{\diff \mSqMiss}
\end{equation}
where $\langle\epsilon_{\mathrm{bkg}}\rangle$ denotes the mean background efficiency of the veto system in this window.
This results in a \emph{detection} sensitivity (at $90\%$ CL) of:
{\small
\begin{multline}
    |U_e|^2 \gtrsim \frac{2.56}{N_K} \frac{\sqrt{\langle\epsilon_{\mathrm{bkg}}\rangle}}{\epsilon_{\mathrm{sig}} (1-\Pdecay{})} \frac{\sqrt{\sigma_{m^2}}}{\mathrm{BR}(K^+ \to \pi^0 e^+ N; |U_e|^2=1)} \\
    \times \sqrt{\frac{\diff N(K^+ \to \pi^0 e^+ \nu_e \gamma)}{\diff \mSqMiss}}
\end{multline}
}
The median \emph{projected exclusion} limit on $|U_e|^2$ from NA62 in the $K^+ \to \pi^0 e^+ N$ channel (valid for any number of quasi-degenerate HNLs) is presented in \cref{fig:sensitivity}, along with the limits set by previous searches using the missing mass technique at KEK~\cite{Yamazaki:1984sj}, PIENU~\cite{Aguilar-Arevalo:2017vlf} and NA62~\cite{NA62:2020mcv}, as well as the so-called seesaw ``bounds'' for both the normal and inverted hierarchy. These lines are the \emph{model dependent} lower bounds on the mixing angle\footnote{For consistency, we have plotted the lower bound on the mixing angle $|U_e|^2$ instead of the commonly used total mixing $U^2$. Our limit is therefore below the usual seesaw bound.} $|U_e|^2 = \sum_{I=1,2} |\Theta_{e I}|^2$ in the type-I seesaw with two Majorana HNLs forming a quasi-Dirac pair. As discussed in \cref{sec:background}, we have assumed $\epsilon_{\mathrm{bkg}} = 10^{-3}$ for in-acceptance photons, which results in an overall background efficiency of $\langle\epsilon_{\mathrm{bkg}}\rangle \approx 1.7\%$ mainly driven by out-of-acceptance photons. If the HNL has a short lifetime (comparable to or smaller than the size of the detector) and a probability of order~$1$ to decay visibly, then the sensitivity to $|U_e|^2$ will be reduced by a factor of $(1-\Pdecay{})^{-1}$ due to fewer events being available for the analysis. But as we saw previously in \cref{sec:signal}, this does \emph{not} suppress the sensitivity to an HNL whose lifetime is suppressed by new \emph{invisible} decays, since those cannot be vetoed even if the HNL decays within the detector. It only matters if non-minimal interactions enhance the visible decay width beyond what is allowed by existing constraints on the mixing angles $|U_{\alpha}|^2$. In particular, this missing mass search has no sensitivity to an HNL which decays promptly to visible particles.

\section{Discussion and outlook}

The black solid line in \cref{fig:sensitivity} represents the sensitivity\footnote{The ($m_N$, $|U_e|^2$) coordinates of the estimated sensitivity curves can be extracted from the file \texttt{figures/sensitivity.tex} in the \LaTeX{} source of the \textsc{arXiv} version of this paper.} achievable with the currently available dataset, corresponding to an effective number of $K^+$ decays of $N_K \approx 3\times 10^{10}$.
The NA62 collaboration is planning to collect an additional dataset in 2021 -- 2024~\cite{Collaboration:2691873}. Assuming no changes to the pre-scaling factors applied to the minimum-bias triggers, this leads to an estimated additional $6\times10^{10}$ effective kaon decays. Considering in addition $K^+\to\pi^0 e^+ N$ decays followed either by the Dalitz decay $\pi^0\to\gamma e^+ e^-$ (which has branching fraction $1.17\%$~\cite{Zyla:2020zbs}) or by a $\pi^0\to\gamma\gamma$ decay with one of the photons converting just upstream of the trigger hodoscope, both of which are recorded by the current di-electron trigger\footnote{The intersection of the di-electron and 1-track triggers used to obtain the current $K^+ \to \pi^0 e^+ \nu/N$ sample is negligible due to the large downscaling factor of the latter.}~\cite{NA62:2759557}, we expect an additional sample corresponding to $5\times 10^{10}$ kaon decays, bringing the total to $1.4\times 10^{11}$ by 2024. The corresponding sensitivity is shown by the black dashed line.

In order for the $K^+ \to \pi^0 e^+ N$ search at NA62 to become truly competitive in the region of interest, and start filling the current gap between $125$ and $\SI{144}{MeV}$, a dedicated trigger line (without the current prescaling factor of $\sim 150$) is required. If NA62 were to implement such a trigger for its 2021 -- 2024 run, it would be able to establish a limit at the level of $|U_e|^2 \approx 10^{-7}$ (represented by the black dotted line) assuming a fully efficient trigger. This is potentially stronger that the exclusion limit set by PS191, when interpreted within the simplest realistic model of HNLs~\cite{Bondarenko:2021cpc}. Finally, any improvement in the rejection of out-of-acceptance photons, for instance through optimized selection or increased veto coverage, would push the sensitivity further down until the missed in-acceptance photons become the leading source of background.

The limits discussed in this paper present little model dependence, as long as the HNL is produced in a flavor-changing kaon decay. The remaining dependence comes from the possibly short lifetime of the HNL, which could in principle induce additional activity in the detector if the HNL decays visibly, resulting in the event being excluded from the present analysis. This limitation does not apply if the new HNL decays responsible for the short lifetime are invisible (\eg{} if the HNL decays into a hidden sector). In order to partially overcome it, it would be interesting to allow for a displaced vertex compatible with the missing momentum. If the HNL lifetime is so short that the displaced vertex cannot be resolved --- as can happen in some non-minimal models, such as the one discussed in Ref.~\cite{Ballett:2019pyw} --- then dedicated searches involving prompt HNL decays will be needed. These searches are, however, inherently model dependent, since they target specific decay channels.%
\newline

\begin{acknowledgments}
The authors are grateful to the NA62 Collaboration for allowing the use of their software code and framework to obtain part of the results presented here.
This project has received funding from the European Research Council (ERC) under the European Union's Horizon 2020 research and innovation programme (GA 336581, 694896) and from the Carlsberg foundation. The work of IT has been supported by ERC-AdG-2015 grant 694896 and by the Swiss National Science Foundation Excellence grant 200020B\underline{ }182864.
We thank the anonymous referees for their thoughtful suggestions.
\end{acknowledgments}

\bibliographystyle{utphys}
\bibliography{refs}

\end{document}